\documentclass[iop]{emulateapj}

\usepackage{txfonts}
\usepackage{graphicx}
\usepackage{natbib}

\newcommand{\ks}{\mbox{km~sec$^{-1}$}}

\def\etal{{\sl et al.}}
\usepackage{natbib}
\shorttitle{Chromospheric bubbles in solar flare models}
\shortauthors{Reid et al.}

\begin{document}

\title{CHROMOSPHERIC BUBBLES IN SOLAR FLARES}
\vskip1.0truecm
\author{
A. Reid$^{1}$, B. Zhigulin$^{1}$, M. Carlsson$^{2,3}$, M. Mathioudakis$^{1}$}
\affil{
1. Astrophysics Research Centre, School of Mathematics and Physics, Queen's University Belfast, BT7~1NN, Northern Ireland, UK; e-mail: aaron.reid@qub.ac.uk\\
2. Rosseland Centre for Solar Physics, University of Oslo, P.O. Box 1029 Blindern, NO-0315 Oslo, Norway \\
3. Institute of Theoretical Astrophysics, University of Oslo, P.O. Box 1029, Blindern, NO-0315, Norway\\
}
\begin{abstract}
We analyse a grid of radiative hydrodynamic  simulations of solar flares to study the energy balance and response of the atmosphere to non-thermal electron beam heating. The appearance of chromospheric bubbles is one of the most notable features that we find in the simulations. These pockets of chromospheric plasma get trapped between the transition region and the lower atmosphere as it is superheated by the particle beam. The chromospheric bubbles are seen in the synthetic spectra, appearing as an additional component to Balmer line profiles with high Doppler velocities as high as 200~kms$^{-1}$. Their signatures are also visible in the wings of Ca II 8542~\AA\ line profiles. These bubbles of chromospheric plasma are driven upward by a wavefront that is induced by the shock of energy deposition, and require a specific heating rate and atmospheric location to manifest.

\end{abstract}

\keywords{Sun: Activity --- Sun: Atmosphere --- Sun: Chromosphere --- Sun: Flares --- Radiative Transfer}

\section{INTRODUCTION}
 \label{Section2}

Solar flares are impulsive events with energies reaching up to 10$^{32}$~erg, caused by the sudden release of free magnetic energy in the corona which is transported into the chromosphere during the impulsive phase of the flare event. The prevailing theory is that this energy transport arises from a propagating beam of accelerated particles driven from the corona (e.g. \cite{Brown1971, Emslie1978, Holman2011}). Alfv\'{e}n waves have also been suggested as an alternative energy transport mechanism \citep{Fletcher2008,Kerr2016,Reep2018}, but this process has not been explored as thoroughly. This manuscript will focus on the primary theory of particle beam heating, specifically electron beams. 

The majority of non-thermal accelerated particles can penetrate into the dense chromosphere, causing intense heating via Coulomb collisions, resulting in X-ray emission via Brehmsstrahlung at the loop foot-points \citep{Hudson1992,Neidig1993,Martinez2011}. This rapid heating causes a pressure gradient which drives a flow upwards, pushing material to greater geometrical heights in a process known as chromospheric evaporation \citep{Neupert1968}. There are 2 main types of chromospheric evaporation \citep{Milligan2006}. Gentle evaporation occurs with lower beam fluxes when thermal heating causes chromospheric expansion via an up-flow of the order of 10 \ks, with little evidence of a corresponding down-flow. Explosive evaporation occurs when the beam flux heats the chromosphere to coronal temperatures, causing the transition region to initially shift to lower geometrical heights. The intense heating cannot be radiated away sufficiently fast and this results in the expansion of the chromosphere with velocities up to 100 \ks. The velocity is strong enough to create a shock, and due to the conservation of momentum, a strong down-flow can also be present, known as chromospheric condensation \citep{Kosovichev1986,Hudson2011}. \cite{Fisher1985} suggest that the flux threshold between gentle and explosive evaporation for a 20~keV low energy cut-off model is $\sim$10$^{10}$ erg cm$^{-2}$ s$^{-1}$. The crossover flux has also been estimated as 2 - 8 x10$^{9}$ erg cm$^{-2}$ s$^{-1}$ using both observations and the F-CHROMA grid of numerical models \citep{Sadykov2018} and has also been estimated with other models \citep{reep} and observations \citep{gomory}. It has also been proposed that the photosphere can be heated directly by electron beams or even via proton beams which can penetrate deeper into the atmosphere \citep{Svestka1970,Machado1978,Aboudurham1986, Prochazka2018}.

The parameters of the non-thermal particle beams can be constrained using  the  X-ray spectrum captured with instruments such as the Reuven Ramaty High Energy Spectroscopic Imager (RHESSI, \cite{Lin2002}) or the Fermi Gamma-ray Space Telescope (FERMI, \cite{Meegan2009}) during the impulsive phase of a flare. Estimates for the cut-off energy, spectral index, and flux of the accelerated electron beam can be inferred from this X-ray spectrum using the Collisional Thick Target Model (CTTM), as demonstrated by \cite{Petrosian2010}.
 
As the majority of the accelerated electrons lose their energy in the chromosphere, this is the part of the solar atmosphere where the majority of the flare radiative output originates \citep{Fletcher2011}. It is therefore imperative to understand how the chromospheric plasma reacts to the dynamics of the magnetic reconnection which mediates the resultant influx of precipitating particles.
In this manuscript, we describe the new phenomenon of chromospheric bubbles which are related to the rate of chromospheric evaporation and particle deposition in the lower solar atmosphere. Section 2 describe the simulations used, while Section 3 characterises the bubbles. Section 4 discusses the overall findings and potential future work.\\
 
 \section{RHD AND RADIATIVE TRANSFER CODES}
  \label{Section3}

The work presented in this paper uses a grid of flare models created with the RADYN code \citep{carlsson1992, carlsson1994, carlsson1995} and are part of the F-CHROMA model archive (https://star.pst.qub.ac.uk/wiki/doku.php/public/solarmodels/start). RADYN solves the equations of radiative hydrodynamics and equation of state (EOS) along a one dimensional atmosphere as described in detail by \cite{Allred2006}. It allows for the introduction of energy via various scenarios, such as direct thermal heating \citep{Reid2017, Prochazka2017}, Alfv\'{e}n wave heating \citep{Kerr2016}, and also via particle beams \citep{Rubiodacosta2016, Prochazka2018} with the modifications created by \cite{Abbett1999} (See also \cite{Allred2005}). The RADYN model atmosphere contains a 6-level hydrogen atom, a 9-level helium atom, and a 6-level calcium atom. It solves the equation of radiative transfer in complete redistribution (CRD) which is a good approximation only for non-resonance lines.

The F-CHROMA flare models utilise a Fokker-Planck type beam \citep{Allred2015, Daou2016}, and use an initial atmosphere with 300 grid points along a VAL3C starting atmosphere. An adaptive grid allows for small-scale dynamic events to be accurately resolved. \cite{Allred2015} compared how the hardness of the applied beam and the low energy cut-off affects the resultant location of the deposition of the non-thermal electrons for both a Fokker-Planck type beam and an Emslie beam \citep{Emslie1978}. The F-CHROMA grid also contain the physics of return currents \citep{Holman2012} which considers how the accelerated electrons produce an electric field which drives a counter-propagating, neutralizing return current. The return current will heat the plasma via Joule heating \citep{vandenOord1990}, but will not largely affect the chromospheric energy deposition. 

RADYN simulates a 10~Mm half loop. The spectral index $\delta$ ranged between 3 - 8 while the low energy cutoff $E_C$ ranged between 10 keV - 25~keV in steps of 5~keV. The total beam fluxes chosen were 3.0e+10 erg cm$^{-2}$, 1.00e+11 erg cm$^{-2}$, and 3.0e+11 erg cm$^{-2}$. This results in a total of 72 models. The simulations were run for a total of 50 seconds, with atmospheric outputs saved every 0.1 seconds. The beam heating is applied for 20 seconds with a triangular temporal profile peaking at T=10 seconds. The parameters selected for the input beam are listed in Table~\ref{table1}.

\begin{table}[!h]
\centering
\begin{tabular}{| c | c | c | c | c | c | c | c | c | c | c | c | c |}
\hline
\textbf{Beam Parameter} & \multicolumn{12}{|c|}{\textbf{Allowed Values}} \\ \hline
\textbf{Flux (erg cm$^{-2}$)} & \multicolumn{4}{|c|}{3.00e10} & \multicolumn{4}{|c|}{1.0e11}  & \multicolumn{4}{|c|}{3.0e11} \\ \hline
\textbf{Low Energy Cut-off (keV)} & \multicolumn{3}{|c|}{10} & \multicolumn{3}{|c|}{15} & \multicolumn{3}{|c|}{20} & \multicolumn{3}{|c|}{25} \\ \hline
\textbf{Spectral Index} & \multicolumn{2}{|c|}{3} & \multicolumn{2}{|c|}{4} & \multicolumn{2}{|c|}{5} & \multicolumn{2}{|c|}{6} & \multicolumn{2}{|c|}{7} & \multicolumn{2}{|c|}{8} \\ \hline
\end{tabular}
\caption{The various allowed input beam parameters for models used in this study.}
\label{table1}
\end{table}

\begin{figure*}[!t]
   \centering
    \includegraphics[width=\textwidth]{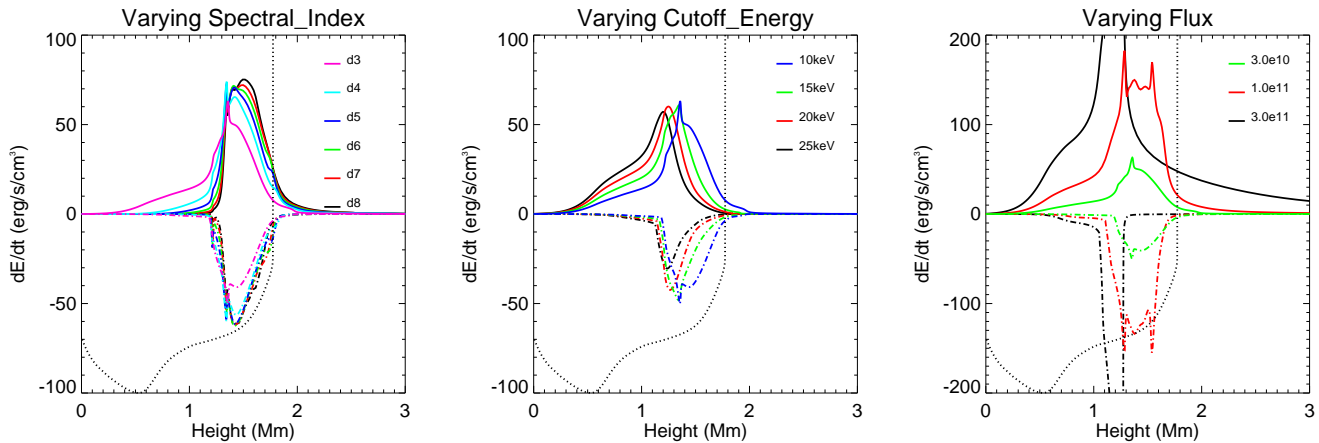}
\caption{Electron beam energy deposition (solid lines) as a function of height. Left: Varying spectral index with a beam flux 3.0e10 erg cm$^{-2}$ and a low energy cutoff of $E_c$=10~keV. Middle: Varying low energy cutoff with a beam flux of 3.0e10 erg cm$^{-2}$ and a spectral index of $\delta$=3. Right:  Varying the beam flux for models with $\delta$=3, $E_c$=10~keV. The dashed-dotted lines indicate the optically thin radiative losses. The over-plotted dotted black line indicates the temperature profile of the starting atmosphere for reference. Positive values are energy input, with negative values output.}
\label{Figure1}
\end{figure*} 



 \section{Chromospheric Bubbles}
 \label{Section5_bubble}

\begin{figure*}[!t]
   \centering
    \includegraphics[height=\textheight]{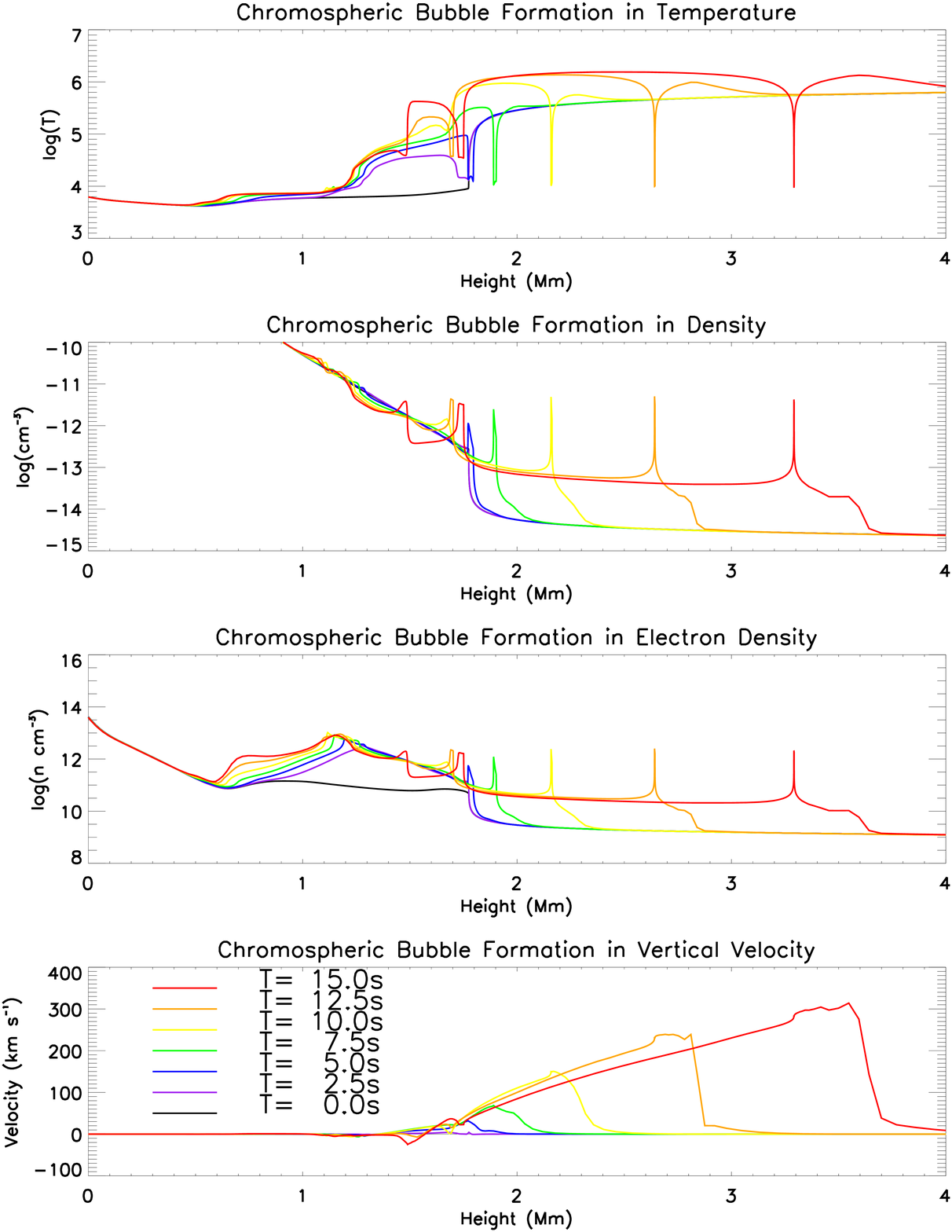}
\caption{The chromospheric bubble propagating through an example atmosphere between T=0s and T=15s of the simulation. First panel: log(Temperature). Second Panel: log(Density). Third Panel: log(Electron Density). Fourth Panel: Velocity (negative = downflow). The colours for all panels correspond to the times denoted in the legend of the fourth panel. The model shown has a flux of 1.0e11 erg cm$^{-2}$, $E_c=15$~keV, and $\delta$=7.}
\label{Figure_bubble}
\end{figure*}

Figure~\ref{Figure1} shows the deposited electron energy and radiative losses from the electron beams  in the simulated grid at a time of T=10s. The left panel shows how varying the spectral index has a relatively small effect on the penetration depth of the beam, while varying the low energy cut-off does. Varying the flux the most significant impact on the temperature increase. If the conditions are just right, the chromosphere will heat up quickly to over 100,000~K. This heating will occur sufficiently deep in the atmosphere to leave a pocket of undisturbed chromosphere sandwiched between the energy deposition and the transition region. As the simulations develop and chromospheric evaporation begins, bubbles of small pockets of chromosphere begin to rise. These bubbles are small regions of chromospheric temperature and increased electron density that also show an increase in mass density relative to the surrounding coronal type atmosphere. These oddities appear mainly in models with high beam fluxes and softer beams. These small pockets of chromospheric plasma also show an increase in gas pressure at the boundaries at either side of the discontinuity. An example of one of these bubbles can be seen in Figure~\ref{Figure_bubble}.

The top panel of Figure~\ref{Figure_bubble} shows clearly some `trapped' plasma with chromospheric temperatures within the corona at T=10-15 seconds. This bubble is formed when the energy is sufficiently high to cause heating over 100,000~K at chromospheric heights. Importantly, this heating must occur sufficiently deep, and be confined so that it does not heat the initial transition region. We essentially have the creation of a chromospheric pocket of plasma between the transition region and the region of the chromosphere heated by the electron beam. To ensure this is not an artefact of the simulations, the same beam was run with no return current. We have also modified the weighting of the grid points in the velocity domain to check whether models which put less emphasis on small scale dynamics in the corona could still resolve the bubbles. In all cases, the bubbles created were identical. However, changing the starting atmosphere will change whether a bubble is identified. This is due to the change in the transition region location with respect to the hydrogen column density. The beam will therefore not penetrate at the same location into a different starting atmosphere, which will not trap a portion of the chromosphere. It may very well be that the bubbles can be created in other starting model atmospheres but with a different set of beam parameters.

In the example shown in Figure~\ref{Figure_bubble}, the initial bubble accelerates upwards into the corona initially due to the pressure gradient caused by the beam heating in the chromosphere. However, a shock forms at T=10 seconds, which highly accelerates the upper atmosphere, including the bubble. By T=15 seconds into the simulation, this velocity is already over 300 \ks. The value of temperature, density, electron density are all stationary within the bubble over time which is essentially the mass motion of the material. Interestingly, due to the gas pressure increasing at the edges of the bubble, the overall width of the bubble decreases over time. For the bubble in Figure~\ref{Figure_bubble}, 45\% of the mass is lost between 10-15 seconds due to redistribution of the bubble mass into the overall wave-front. This compression can be most easily seen from T=5 seconds to T=10 seconds in Figure~\ref{Figure_bubble}.

Individual well defined bubbles were identified by investigating the temperature of each atmosphere. The location of the transition region was defined to be the lowest point in the atmosphere where the temperature gradient exceeds 5,000~K between grid cells, while having a temperature value of at least 50,000~K. Any point above the transition region which has a temperature below 40,000~K was deemed to be a part of the bubble and logged. This resulted in 21/72 atmospheres with bubbles. Of these 21, all had beam fluxes above 1.0e11 erg cm$^{-2}$, and low energy cut-off values above 15~keV. The most common value of low energy cut-off being $E_c = 15$~keV.  This value provides the optimal penetration depth to efficiently trap a portion of the chromosphere while accelerating it upwards (see Figure~\ref{Figure1}). The $E_c$=10~keV models mainly heated the upper chromosphere and some of the transition region, and did not lead to the creation of bubbles. The $E_c$=25~keV models only resulted in bubbles for the highest spectral indices ($\delta > 5$) with the highest levels of beam flux (3.0e11 erg cm$^{-2}$). This is due to the extremely high energy density that needs to be applied to the lower chromosphere in order to heat it sufficiently and create a bubble. Harder beams tend to have a larger spread of energy into the denser portions of the photosphere and lower chromosphere.

The bubble lifetimes can be as short as a few seconds or survive to the end of the 50 second simulations. Once bubbles reach the top of the 10~Mm half loop structure, they are reflected due to the top boundary condition. Any physical changes in the atmosphere post-reflection are not considered as realistic as this would imply symmetrical bubbles being created on either side of the loop. The lifetime of a bubble in the simulation ends when the simulation can no longer resolve it due to the minimum resolution between grid points ($\sim$1 km). All bubbles will exponentially shrink with time due to external forces pushing inwards. Once a bubble pops, the simulation then redistributes the grid points, inducing some numerical artefacts which appear to look like waves. This artefact will be damped, and the atmosphere will settle to a new equilibrium. This means that the lifetime of a bubble can not be accurately determined from these simulations. It remains unclear as to what happens to a bubble once it shrinks below 1~km in width.

The upward movement of the bubbles will always have a roughly constant acceleration in line with the acceleration of the upwardly moving wavefront from the chromospheric shock. This can range from 4 - 77~km s$^{-2}$, depending on the formation height and point along the wavefront the bubble first appears, as well as the amplitude of the wavefront itself, as the bubble essentially `rides' along this wavefront.

Some beams will result in multiple bubbles. Secondary bubbles will generally form only after the initial bubble has progressed into the corona (usually after 10 seconds), and while beam heating is still being applied. The secondary bubbles are usually formed lower in the atmosphere, due to the atmospheric changes that have occurred from the previous heating event. As such, they will move into the corona with a smaller acceleration as they are formed in denser media causing less upward propagation. Tertiary bubbles also exist in 3 of the most extreme simulations, with 3.0e11 erg cm$^{-2}$ beam fluxes, $\delta > 6$, and $E_c$=15 keV.

\begin{figure*}[!t]
   \centering
    \includegraphics[height=0.45\textheight]{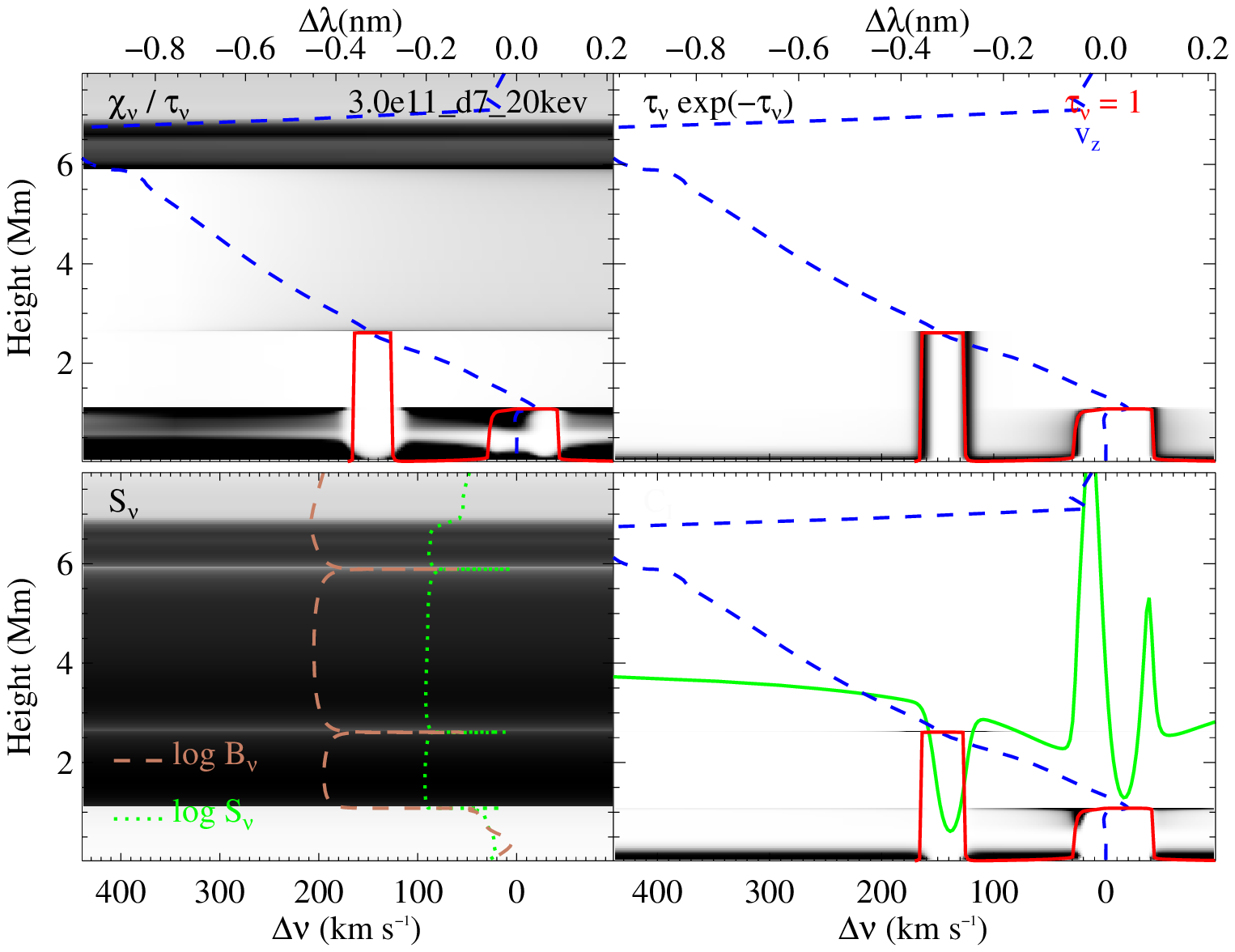}
    \includegraphics[height=0.45\textheight]{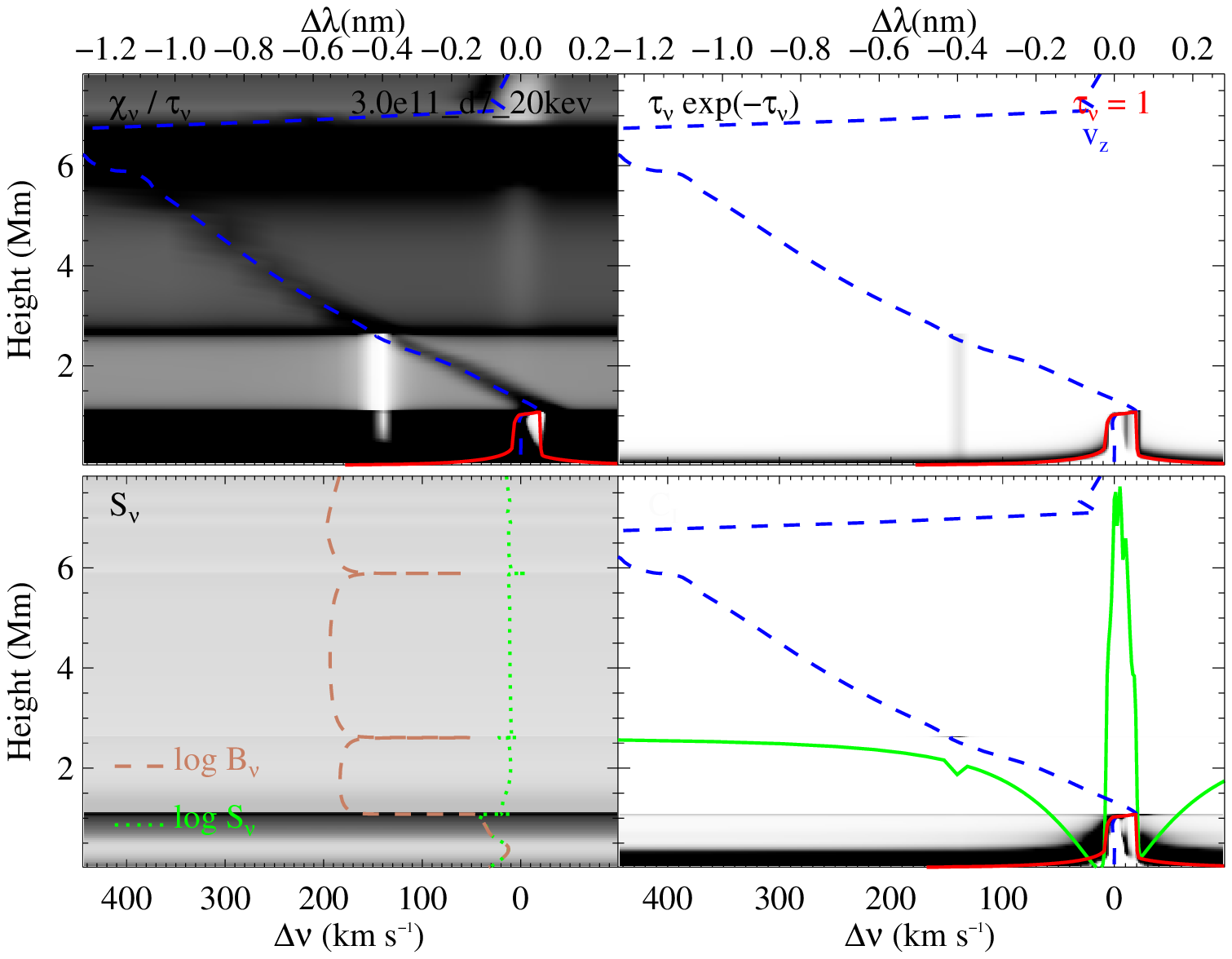}
\caption{Top 4-panels:  Contribution functions for H$\alpha$ at T=21s for a beam of 3.0e11 erg cm$^{-2}$, $\delta$=7, $E_c$=20~keV. The green line shows the line profile. The blue dashed lines are the vertical velocity components where negative values correspond to upflows.  The top left panel shows the opacity divided by the optical depth. The top right panel show the optical depth multiplied by the negative exponent of the optical depth, while the bottom left panel indicate the source function. The bottom right panels highlight the contribution function (black areas correspond to strong contribution). Bottom 4-panels: The corresponding plots for Ca II 8542~\AA.}
\label{Figure_contrib}
\end{figure*}

In order to estimate the observational signatures of the bubbles, we carried out radiative transfer calculations in H$\alpha$ and Ca II 8542~\AA. Figure~\ref{Figure_contrib} shows the line contribution functions along with the corresponding line profiles (shown in green) for H$\alpha$ and Ca II 8542~\AA\ for a 3.0e11 erg cm$^{-2}$ beam flux with $\delta$=7 and $E_c$=20~keV. The contribution functions shown are described in detail in \cite{Carlsson1997}.

 A primary bubble exists near the peak of the velocity, at around 6~Mm. This can be seen in the bottom left panels of Figure~\ref{Figure_contrib} where the source function (green line) and Planck function (brown line) are shown. The bubble is optically thin with the $\tau$=1 location at the corresponding Doppler wavelength being in the lower photosphere, with no signs of signal in the resultant intensity. However, the secondary bubble (at 2.6~Mm) does show significant H$\alpha$ and some Ca II 8542~\AA\ signal at a Doppler velocity of 140 \ks. The chromosphere at this location, combined with a strong velocity gradient, is sufficiently dense to cause a large increase in the optical depth ($\tau_\nu$). The sudden change into a much denser medium increases the opacity proportionately, as can be seen at a height of 2.6~Mm in the top left panels of both sub-plots of Figure~\ref{Figure_contrib}.

In this instance, the bubble appears optically thick in the H$\alpha$ line profile, showing strong absorption features as it appears like a quiet-Sun piece of chromosphere. The Ca II 8542~\AA\ spectra however show the bubble as optically thin, with only a minor change in the line profile. These bubbles have also been recently shown to be dominant in the Lyman $\alpha$ line of hydrogen by \cite{Brown2018}.




 \section{Conclusions}
We have utilised the publicly available F-CHROMA flare model archive hosted by Queen's University Belfast to study the response of the solar atmosphere to electron beam heating. Certain beam parameters lead to the generation of chromospheric bubbles propagating into the corona. These bubbles are confined regions of chromospheric plasma trapped between the transition region and the lower chromosphere that has been heated by the electron beam. The resultant shock wavefront accelerates the bubble upward into the corona. The bubbles will compress over time, but can be accelerated to Doppler velocities of up to 500 \ks. These bubbles will eventually disappear as they compress, releasing an artificial numerical artefact throughout the atmosphere which we do not consider realistic and solely due to the redistribution of the grid points. The bubbles are only formed if sufficient heating occurs in the chromosphere without affecting the transition region, and so require optimal electron beam parameters to manifest. A small number of 1.0e12 erg cm$^{-2}$ flux beam models were also created, with all showing evidence of bubbles. Only those models which show explosive evaporation contain bubbles. The bubbles appear strong in the H$\alpha$ line up to 10~\AA\ from line core. The Ca II 8542~\AA\ line profiles also show an optically thin remnant of the bubbles. Not all bubbles appear as optically thick in H$\alpha$, and indeed some show no noticeable effects on the calculated line profiles. As such, only a small portion of beams will produce these bubbles, and even fewer can be observationally detected. The observational signatures of the bubbles could be detected with an instrument that covers a broad spectral range around chromospheric lines and  provides a good signal to noise ratio. To our knowledge, no observations currently exist for solar flares which show observational evidence of the bubble. However, these up-flowing bubbles may have been observed previously in a flaring dMe star \citep{Gunn1994} with similar velocities, only assumed to be high-velocity evaporation instead. The large flux of the flare observed aligns with the theory presented in this manuscript that stronger fluxes are more likely to produce these bubbles.
 
 \begin{acknowledgements}

The research leading to these results has received funding from the European Community\'s Seventh Framework Programme (FP7/2007-2013) under grant agreement no. 606862 (F-CHROMA), and from the Research Council of Norway through the Programme for Supercomputing.  

\end{acknowledgements}

\end{document}